\documentstyle[12pt]{article}
\def\one{1\hskip-.37em 1}
\def\ir{{\rm I}\hskip-.2em{\rm R}}
\def\half{\textstyle{\frac{1}{2}}}

\def\ra{\rightarrow}
\def\D{{\cal D}}
\def\H{{\cal H}}
\def\tint{{\textstyle \int}}
\def\d{\partial}
\def\o{\overline}
\def\b{\begin{eqnarray*}}     
\def\e{\end{eqnarray*}}       
\def\bn{\begin{eqnarray}}     
\def\en{\end{eqnarray}}       
\def\<{\langle}
\def\>{\rangle}
\def\{{\lbrace}
\def\}{\rbrace}
\title{Metrical Quantization\footnote{Presented at the workshop on Quantum
Future, Przesieka, Poland, September, 1997}}
\author{John R.~Klauder\\
Departments of Physics and Mathematics\\
University of Florida\\
Gainesville, Fl  32611}
\begin{document}
\maketitle

\begin{abstract}
Canonical quantization may be approached from several different starting 
points. The usual approaches involve promotion of $c$-numbers to $q$-numbers, 
or path integral constructs, each of which generally succeeds only in 
Cartesian coordinates. All quantization schemes that lead to Hilbert 
space vectors and Weyl operators---even those that eschew Cartesian 
coordinates---implicitly contain a metric on a flat phase space. This 
feature is demonstrated by studying the classical and quantum 
``aggregations'', namely, the set of all facts and properties resident
 in all classical and quantum theories, respectively. Metrical quantization
 is an approach that elevates the flat phase space metric inherent in any 
canonical quantization to the level of a postulate. Far from being an 
unwanted structure, the flat phase space metric carries essential physical 
information. It is shown how the metric, when employed within a 
continuous-time regularization scheme, gives rise to an unambiguous 
quantization procedure that automatically leads to a canonical 
coherent state representation. Although attention in this paper is 
confined to canonical quantization we note that alternative, nonflat 
metrics may also be used, and they generally give rise to qualitatively 
different, noncanonical quantization schemes.
\end{abstract}
\section{Introduction}
Quantization, like any other procedure, lends itself to an axiomatization. 
As discussed shortly, there are such procedures that characterize the 
usual quantization proposals of Heisenberg, Schr\"odinger, and Feynman. 
Hidden in these procedures is an often unstated assumption, namely, that 
the coordinates in which the very quantization rules are laid down must be 
chosen to be Cartesian whenever a canonical quantization is sought. This 
procedural step is so ingrained and automatic that it is often overlooked 
or ignored for what it really is, namely, an essential assumption in the 
given procedure. In this paper we briefly review postulates of the usual 
quantization procedures and introduce yet another procedure we refer to as 
metrical quantization. 

Let us start with a brief review of classical mechanics.
\subsection{Classical mechanics}
Consider a phase space $\cal M$ for a single degree of freedom which is 
two dimensional. As a symplectic manifold the space $\cal M$ is endowed 
with a symplectic two form $\omega$, which is nondegenerate and closed, 
$d\omega=0$. Darboux's Theorem assures us that local coordinates $p$ and 
$q$ exist such that $\omega=dp\wedge dq$ in the given coordinates. Such 
coordinates are referred to as canonical coordinates, and any coordinate 
transformation with a unit Jacobian leads from one set of canonical 
coordinates to another set of canonical coordinates. Indeed, if $r$ and 
$s$ denote another pair of canonical coordinates, then it follows that 
$rds=pdq+dF(s,q)$ for some generator $F$. The new coordinates are canonical 
since the exterior derivative of both sides of this relation yields 
$dr\wedge ds=dp\wedge dq=\omega$.

Besides the kinematical aspects of the classical theory of mechanics, 
dynamics arises with the introduction of a distinguished scalar, the 
Hamiltonian $H$, or as expressed in the original canonical coordinates, 
the function $H(p,q)$. By a scalar we mean that ${\o H}(r,s)\equiv 
H(p(r,s),q(r,s))=H(p,q)$, an equation which indicates how $H$ transforms 
under (canonical) coordinate transformations. Finally, classical dynamics 
may be introduced as the stationary paths of a distinguished action 
functional given in coordinate form by
  \bn I=\tint[p{\dot q}+{\dot G}(p,q)-H(p,q)]\,dt\;,  \en
subject to variations that hold both $p(t)$ and $q(t)$ fixed at the 
initial time $t=0$ and the final time $t=T$. The resultant equations are 
independent of the gauge function $G$, and are given by
  \bn &&{\dot q}=\d H(p,q)/\d p\;,  \nonumber\\
      &&{\dot p}=-\d H(p,q)/\d q\;.   \en
Note that the exterior derivative of the one form $pdq+dG(p,q)$ that 
appears in the action functional leads to $d[pdq+dG(p,q)]=dp\wedge 
dq=\omega$. In this way the symplectic structure enters the dynamics. 

Lastly, we observe that the dynamical equations of motion may also be 
given a Poisson bracket structure. In particular, if 
  \bn  \{A,B\}\equiv \frac{\d A}{\d q}\frac{\d B}{\d p}-
\frac{\d A}{\d p}\frac{\d B}{\d q}\;,  \en
then it follows that
  \bn &&{\dot q}=\{q,H(p,q)\}\;,  \nonumber \\
    &&{\dot p}=\{p,H(p,q)\}\;,  \en
and for a general function $W(p,q)$ it follows that
  \bn  {\dot W}(p,q)=\{W(p,q),H(p,q)\}\;.  \en
Thus, since $\{B,A\}=-\{A,B\}$, we observe that
  \bn  {\dot H}(p,q)=\{H(p,q),H(p,q)\}=0\;, \en
and therefore $H(p,q)=E$, which is a constant of the motion usually 
identified with the energy. 
\section{The Classical Mechanics Aggregation}
Let us collect all the concepts and formulas appropriate to classical 
mechanics, a few of which have been indicated above, in one place, and 
let us refer to that as the {\it classical mechanics aggregation}. For 
example, the classical mechanics aggregation would include the set of all 
canonical coordinates, the set of all Hamiltonians each of which is 
expressed in all possible canonical coordinates, the rules for dynamical 
evolution, and indeed the set of all solutions of the dynamical equations 
of motion for each Hamiltonian expressed in all possible canonical 
coordinates. Also in the classical mechanics aggregation would be the 
formulation of classical mechanics expressed in differential geometric 
form, i.e., as coordinate-free expressions and operations that effect 
the Poisson brackets, etc. Evidently, the classical mechanics aggregation 
contains all that is known and, implicitly, all that is knowable about 
classical mechanics!

Let us develop an analogous aggregation appropriate to quantum mechanics.
\section{The Quantum Mechanics Aggregation}
There are a number of standard ideas and equations that enter into the 
formulation of quantum mechanics irrespective of the particular details of 
the system being quantized, and, for purposes of illustration, let us 
focus on systems with just one degree of freedom. We have in mind, for 
example, a Hilbert space composed of complex, square integrable functions 
over the real line, namely the space $L^2(\ir)$, or a Hilbert space 
composed of square summable sequences, namely the space $l^2$, etc. 
Operators arise in the form of functions of position and derivatives with 
respect to position, or functions of momentum and derivatives with respect 
to momentum, or semi-infinite square matrices, etc. Probability amplitudes 
occur in the form of inner products of two Hilbert space vectors, or more 
generally, matrix elements of an operator in the form of an inner product 
involving two vectors with an operator standing between them. Many of these 
concepts can be formulated in a coordinate-free language in terms of an 
abstract Hilbert space formulation and an abstract operator language as 
well. These elements form the arena in which quantum mechanics takes 
place. Quantum mechanics is also distinguished by equivalent sets of 
rules for the introduction of dynamics. For example, there is the 
abstract Schr\"odinger equation giving the time derivative of the state 
vector as the action of the Hamiltonian operator on the state vector, 
apart from suitable constants $(i\hbar)$. Alternatively, there is the 
Heisenberg equation of motion which equates the time derivative of an 
operator in the Heisenberg picture to the commutator of the operator 
with the Hamiltonian, again up to the same constants. Additionally, we 
mention the Feynman representation of the propagator as a path integral, 
a representation which in fact is a direct consequence of the abstract 
vector and operator language, or alternatively, a consequence of the 
Schr\"odinger equation and its solution for a suitable boundary condition. 

We may also mention distinguished operator sets such as the Heisenberg 
canonical operators $P$ and $Q$ which, either abstractly or in a concrete 
realization, satisfy the fundamental commutation relation $[Q,P]=i\hbar\one$. 
If these operators are self adjoint then we may also consider the Weyl 
operators $U[p,q]\equiv\exp[i(pQ-qP)/\hbar]$ for all real $p$ and $q$. 
Armed with such operators and an arbitrary normalized vector in the Hilbert 
space $|\eta\>$, we may consider the canonical coherent states
  \bn  |p,q\>\equiv|p,q;\eta\>\equiv e^{i(pQ-qP)/\hbar}|\eta\>\;.  \en
It is but a simple exercise to show that
 \bn  \tint |p,q\>\<p,q|\,dp dq/2\pi\hbar=\one  \en
for any choice of the fiducial vector $|\eta\>$. Thus, coherent states, 
the representations of Hilbert space they induce, etc., are all implicitly 
contained within the {\it quantum mechanical aggregation}. Unitary 
transformations that map one form of Hilbert space vectors and one form 
of operators into another form are all part of the quantum mechanical 
aggregation. In short, everything kinematical and dynamical that one 
could think of belonging to Hilbert space, operator theory, quantum 
mechanics, etc., everything known and, implicitly, everything knowable 
about quantum mechanics is contained in the quantum mechanical aggregation. 
 
Now let us try to build a bridge between the classical mechanical aggregation 
and the quantum mechanical aggregation. 
\section{Conventional Quantization}
The act of quantization is designed to connect the principal entities in 
the classical mechanical aggregation with the appropriate entities in the 
quantum mechanical aggregation, in some cases in a one-to-one fashion, but 
in other cases in a many-one fashion. It is the genius of Heisenberg and 
Schr\"odinger that they were able to guess several basic concepts and 
quantities lying in the quantum mechanical aggregation and use these 
few ideas as stepping stones in order to construct a bridge between the 
classical and the quantum worlds. Feynman used a different set of 
concepts and quantities to select his stepping stones between these two 
worlds. In modern parlance, we could call these stepping stones 
``postulates'' (or at the very least ``assumptions''). 
\subsubsection*{Heisenberg quantization}
In the case of Heisenberg quantization, we may cast the postulates in 
the form (for postulate 1. see below):\vskip.2cm
\hskip1cm 2. Introduce matrices $Q=\{Q_{mn}\}$ and $P=\{P_{mn}\}$, 
where $m,n\in\{1,2,3,\ldots\}$, that satisfy $[Q,P]_{mn}\equiv
\Sigma_p( Q_{mp}P_{pn}-P_{mp}Q_{pn})=i\hbar\delta_{mn}$.\vskip.02cm
\hskip1cm 3. Build a Hamiltonian matrix $H=\{H_{mn}\}$ as a function 
(e.g., polynomial) of the matrices, $H_{mn}=H(P,Q)_{mn}$, that is the 
same function as the classical Hamiltonian $H(p,q)$. (In so doing there 
may be operator ordering ambiguities which this prescription cannot 
resolve; choose an ordering that leads to a Hermitian operator.)\vskip.02cm
\hskip1cm 4. Introduce the equation of motion $i\hbar{\dot X}_{mn}=
[X,H]_{mn}$ for the elements of a general matrix $X=\{X_{mn}\}$.\hskip.3cm$
\Box$\vskip.2cm
Along with these postulates comes the implicit task of solving the called 
for equations of motion subject to suitable operator-valued boundary 
conditions.
Once the several steps are accomplished, a general path has opened up 
as to how a given system is to be taken from its classical version to 
its quantum version. 

Accepting these postulates, it becomes clear how the general classical 
system is to be connected with the general quantum system {\it apart 
from one postulate that we have neglected and which was not immediately 
obvious to the founding fathers}. The question arises as to {\it exactly 
which choice of canonical coordinates are to be used when promoting the 
classical canonical variables to quantum canonical variables}. After the 
principal paper on quantization \cite{3men}, it subsequently became clear 
to Heisenberg that {\it it is necessary to make this promotion from 
$c$-number to $q$-number variables only in Cartesian coordinates}. Thus 
there is implicitly another postulate \cite{dir}:\vskip.2cm
\hskip 1cm 1. Express the classical kinematical variables $p$ and $q$ in 
Cartesian coordinates prior to promoting them to matrices $\{P_{mn}\}$ 
and $\{Q_{mn}\}$, respectively.
\vskip.2cm
We will present a rationale for this postulate below.
\subsubsection*{Schr\"odinger quantization}
The postulates for Schr\"odinger's formulation of quantization may be 
given in the following form \cite{sch}:\vskip.2cm
\hskip 1cm 1. Express the classical kinematical variables $p$ and $q$ in 
Cartesian coordinates. \vskip.02cm
\hskip1cm 2. Promote the classical momentum $p$ to the differential 
operator $-i\hbar(\d/\d x)$ and the classical coordinate $q$ to the 
multiplication operator $x$, a choice that evidently satisfies the 
commutation relation $[x,-i\hbar(\d/\d x)]=i\hbar$.\vskip.02cm
\hskip1cm 3. Define the Hamiltonian operator ${\cal H}$ as the classical 
Hamiltonian with the momentum variable $p$ replaced by the operator 
$-i\hbar(\d/\d x)$ and the coordinate variable $q$ replaced by the 
operator $x$. (In so doing there may be operator ordering ambiguities 
which this prescription cannot resolve; choose an ordering that leads to a 
Hermitian operator.)\vskip.02cm
\hskip1cm 4. For $\psi(x)$ a complex, square integrable functions of $x$, 
introduce the dynamical equation $i\hbar{\dot\psi}={\cal H}\psi$.
\hskip.3cm$\Box$\vskip.2cm
Implicit with these postulates is the instruction to solve the 
Schr\"odinger equation for a dense set of initial conditions and 
a large class of Hamiltonian operators, and in that way help to build 
up the essentials of the quantum mechanical aggregation.

It is interesting to note that Schr\"odinger himself soon became aware 
of the fact that his procedure generally works only in Cartesian 
coordinates. \subsubsection*{Feynman quantization}
Feynman's formulation of quantization focuses on the solution to the 
Schr\"o-dinger equation and postulates that the propagator, an integral 
kernel that maps the wave function (generally in the Schr\"odinger 
representation) at one time to the wave function at a later time, may be 
given by means of a path integral expression \cite{fey}.
On the surface, it would seem that the (phase space) path integral, using 
only concepts from classical mechanics, would seem to get around the need 
for Cartesian coordinates; as we shall see that is not the case. As 
postulates for a path integral quantization scheme we have:\vskip.2cm
\hskip 1cm 1. Express the classical kinematical variables $p$ and $q$ 
in Cartesian coordinates.\vskip.02cm
\hskip1cm 2. Given that $|q,t\>$, where $Q(t)|q,t\>=q|q,t\>$, denote 
sharp position eigenstates, write the transition matrix element in the 
form of a path integral as
\bn && \<q'',T|q',0\>={\cal M}\int\exp\{(i/\hbar)\tint[p{\dot q}-
H(p,q)]\,dt\}\,\D p\,\D q\;. \en \vskip.02cm
\hskip1cm 3. Recognize that the formal path integral of Step 2. is 
{\it effectively undefined} and replace it by a {\it regularized} form 
of path integral, namely,
\bn &&\hskip-2.7cm\<q'',T|q',0\>=\lim_{N\ra\infty}M_N\int\exp\{(i/\hbar)
\Sigma_{l=0}^N[p_{l+1/2}(q_{l+1}-q_l)\nonumber\\  &&-\epsilon H(p_{l+1/2},
(q_{l+1}+q_l)/2)]\}\,\Pi_{l=0}^Ndp_{l+1/2}\,\Pi_{l=1}^Ndq_l\;, \en
where $q_{N+1}=q''$, $q_0=q'$, $M_N=(2\pi\hbar)^{-(N+1)}$, and $\epsilon 
=T/(N+1)$.\hskip.3cm$\Box$\vskip.2cm
Implicit in the latter expression is a Weyl ordering choice to resolve 
any operator ordering ambiguities. Observe that the naive lattice 
formulation of the classical action leads to correct quantum mechanical 
results, generally speaking, only in Cartesian coordinates. Although the 
formal phase space path integral of postulate 2.~appears superficially 
to be covariant under canonical coordinate transformations, it would be 
incorrect to conclude that was the case inasmuch as it would imply that 
the spectrum of diverse physical systems would be identical. In contrast, 
the naive lattice prescription applies only to Cartesian coordinates, 
the same family of coordinates singled out in the first postulate of 
each of the previous quantization schemes.
\subsection{Elements of the quantum mechanical aggregation}
Traditional quantization---be it Heisenberg, Schr\"odinger, or 
Feynman---leads invariably to a Hilbert space (or a particular 
representation thereof), and to canonical operators (or particular 
representations thereof). For physical reasons we restrict attention to 
that subclass of systems wherein the canonical operators are self 
adjoint and obey not only the Heisenberg commutation relations but also 
the more stringent Weyl form of the commutation relations. In particular, 
we assert that a byproduct of any conventional quantization scheme---and 
even some nonconventional quantization schemes such as geometric 
quantization or deformation quantization---is to lead to (normalized) 
Hilbert space vectors, say $|\eta\>$, and a family of unitary Weyl 
operators $U[p,q]\equiv\exp[i(pQ-qP)/\hbar]$, $(p,q)\in \ir^2$, that 
obey the standard Weyl commutation relation. These expressions lead 
directly to a set of coherent states each of the form $|p,q\>\equiv 
U[p,q]|\eta\>$. Given such conventional quantities lying in the 
quantum mechanical aggregation, and minimal domain assumptions, we 
first build the one form
   \bn \theta(p,q)\equiv i\hbar\<p,q|d|p,q\>=\half(p\,dq-q\,dp)+\<P\>
\,dq-\<Q\>\,dp\;,\en
where $\<(\,\cdot\,)\>\equiv\<\eta|(\,\cdot\,)|\eta\>$, and which is 
recognized as a natural candidate for the classical symplectic 
potential for a general $|\eta\>$. Indeed, $d\theta=dp\wedge dq=\omega$ 
holds for a general $|\eta\>$. As a second quantity of interest, we 
build the Fubini-Study metric
  \bn \hskip-.9cm&&d\sigma^2(p,q)\equiv2\hbar^2[\|d|p,q\>\|^2-|\<p,q|d
|p,q\>|^2]\nonumber\\
      &&\hskip-.5cm=2\<(\Delta Q)^2\>dp^2+2\<(\Delta P)(\Delta Q)+
(\Delta Q)(\Delta P)\>dpdq+2\<(\Delta P)^2\>dq^2. \en
Here  $\Delta Q\equiv Q-\<Q\>$, etc. The latter expression given above 
holds for a general vector $|\eta\>$. Observe well, for a general 
$|\eta\>$, that this phase space metric is always {\it flat} because all 
the metric coefficients are constants. Stated otherwise, for a general 
$|\eta\>$, the Fubini-Study metric invariably describes a {\it flat 
phase space}, here expressed in (almost) Cartesian coordinates thanks 
to the use of canonical group coordinates for the Weyl group. For 
``physical'' vectors $|\eta\>$, defined such that $\<(\Delta Q)^2+
(\Delta P)^2\>={\rm o}(\hbar^0)$, it follows that the phase space 
metric is a {\it quantum property} and it vanishes in the limit 
$\hbar\ra0$; indeed, if $|\eta\>$ is chosen as the ground state of 
a harmonic oscillator with unit angular frequency, then $d\sigma^2=
\hbar(dp^2+dq^2)$. One may of course change the coordinates, e.g., 
introduce $r=r(p,q)$ and $s=s(p,q)$; this may change the form of the 
metric coefficients for $d\sigma^2$, but it will not alter the fact 
that the underlying phase space is still flat. 

We conclude these remarks by emphasizing that inherent in any canonical 
quantization scheme is the implicit assumption of a flat phase space 
which can carry globally defined Cartesian coordinates. These properties 
automatically lie within the quantum mechanical aggregation for any 
quantization scheme that leads to Hilbert space vectors and canonical 
operators!
\section{Metrical Quantization}
We define metrical quantization by the following set of postulates:
\vskip.2cm
\hskip1cm 1. Assign to classical phase space a flat space metric 
$d\sigma^2$, and choose Cartesian coordinates in such a way that
          \bn d\sigma^2(p,q)=\hbar(dp^2+dq^2)\;. \en\vskip.02cm
\hskip1cm 2. Introduce the regularized phase-space path integral, 
which explicitly uses the phase space metric, and is formally given by 
   \bn &&\hskip-1cm K(p'',q'',T;p',q',0)=\lim_{\nu\ra\infty}
{\cal N}_\nu\int\exp\{(i/\hbar)\tint[(p{\dot q}-q{\dot p})/2-h(p,q)]
\,dt\}\nonumber\\
&&\hskip5cm\times\exp\{-(1/2\nu)\tint[{\dot p}^2+{\dot q}^2]\,dt\}\,
\D p\,\D q  \en
and more precisely given by
   \bn &&\hskip-1cm K(p'',q'',T;p'q',0)\nonumber\\
      &&\hskip-.8cm=\lim_{\nu\ra\infty}2\pi\hbar e^{\nu T/2}\int\exp
\{(i/\hbar)\tint[(pdq-qdp)/2-h(p,q)\,dt]\}\,d\mu_W^\nu(p,q)\,,  \en
where $\mu_W^\nu$ denotes a Wiener measure for two-dimensional 
Brownian motion on the plane expressed in Cartesian coordinates, and 
where $\nu$ denotes the diffusion constant. Finally, we observe that a
s a positive-definite function, it follows from the GNS (Gel'fand, 
Naimark, Segal) Theorem that 
  \bn &&\hskip-.5cm K(p'',q'',T;p',q',0)\equiv\<p'',q''|\,e^{-i\H 
T/\hbar}\,|p',q'\>\;,\\
  &&\hskip1cm|p,q\>\equiv e^{i[(pQ-qP)/\hbar]}|0\>\;,\hskip1cm 
[Q,P]=i\hbar\one\;,\\
&&\hskip1cm(Q+iP)|0\>=0\;,\hskip1cm\<0|0\>=1\;, \\
  &&\hskip1cm \H\equiv\tint h(p,q)\,|p,q\>\<p,q|\,dp\,dq/2\pi
\hbar\;.  \en
All these things follow from positive-definiteness, and the implication 
is that the Wiener measure regularized phase-space path integral 
automatically gives rise to the propagator expressed in a 
coherent-state representation.\hskip.3cm$\Box$\vskip.2cm
The canonical quantization formulation given above has raised the 
metric on a flat phase space to the level of a postulate. The 
assumption that the given coordinates are indeed Cartesian is by 
no means an arbitrary one. There is, in fact, a great deal of 
physics in the statement that certain coordinates are Cartesian. 
In the present case, we can read that physics straight out of (19) 
which relates the classical Hamiltonian $h(p,q)$ to the quantum 
Hamiltonian operator $\H$. The given integral representation is 
in fact equivalent to antinormal ordering, i.e., the monomial 
$(q+ip)^k(q-ip)^l$ is quantized as the operator $(Q+iP)^k(Q-iP)^l$ 
for all nonnegative integers $k$ and $l$. Thus, for example, in 
these coordinates, the $c$-number expression $p^2+q^2+q^4$ is 
quantized as $P^2+Q^2+Q^4+{\rm O}(\hbar)$. Of course, the latter 
term can be made explicit; here we are only interested in the fact 
that the leading terms [${\rm O}(\hbar^0)$] of the quantum 
Hamiltonian operator are exactly those as given by the classical 
Hamiltonian. This connection may seem evident but that is far 
from the case.

Observe that expressed in terms of the Brownian motion regularization, 
and when we define the stochastic integral $\tint p\,dq$ via a 
(midpoint) Stratonovich prescription (as we are free to do in 
Cartesian coordinates), the procedure of metrical quantization 
is actually {\it covariant} under canonical coordinate transformations. 
As noted earlier, such a transformation is determined by the expression 
$r\,ds=p\,dq+dF(s,q)$ in the classical theory, and, thanks to the S
tratonovich prescription, also in the quantum theory where the paths 
$p$ and $q$ are Brownian motion paths. The function $h$ transforms 
as a scalar, and therefore ${\o h}(r,s)\equiv h(p(r,s),q(r,s))=h(p,q)$. 
Lastly, we transform the Wiener measure which still describes Brownian 
motion on a flat two-dimensional plane, but now, generally speaking, in 
curvilinear coordinates. After the change of coordinates the propagator 
reads
  \bn  &&\hskip-.8cm{\o K}(r'',s'',T;r',s',0)\nonumber\\
&&\hskip-.7cm=\lim_{\nu\ra\infty}2\pi\hbar e^{\nu T/2}\int\exp\{(i/\hbar)
\tint[
r\,ds+dG(r,s)-{\o h}(r,s)\,dt]\}\,d{\o \mu}_W^\nu(r,s). \en
Here, $dG$ denotes a total differential, which amounts to nothing more 
than a phase change of the coherent states, and ${\o \mu}_W^\nu$ denotes 
Brownian measure on the flat two-dimensional plane expressed now in 
curvilinear coordinates rather than Cartesian coordinates. In this 
case the connection of the classical and quantum Hamiltonians is given by
  \bn\H=\tint {\o h}(r,s)\,|p(r,s),q(r,s)\>\<p(r,s),q(r,s)|\,dr\,ds/2
\pi\hbar\;. \en
Observe in this coordinate change that the coherent states have 
remained unchanged (only their names have changed) and, as a 
consequence, the Hamiltonian operator $\H$ is {\it absolutely 
unchanged} even though its $c$-number counterpart (symbol) is now 
expressed by ${\o h}(r,s)$. In other words, the leading [${\rm O}
(\hbar^0)$] dependence of ${\o h}$ and $\H$ are no longer identical. 
As we have stressed elsewhere \cite{namiki}, the {\it physical 
significance} of the {\it mathematical expression} for a given 
classical quantity is {\it encoded} into the specific coordinate 
form of the auxiliary metric $d\sigma^2$; for example, if the 
metric is expressed in Cartesian coordinates, then the physical 
meaning of the classical Hamiltonian is that directly given by 
its coordinate form, as has been illustrated above by the anharmonic 
oscillator. Since quantization deals, for example, with the highly 
physical energy spectral values, it is {\it manditory} that the 
mathematical expression for the Hamiltonian somehow ``know'' to which 
physical system it belongs. It is the role of the metric and the very 
form of the metric coefficients themselves to keep track of just what 
physical quantity is represented by any given mathematical expression.  
And that very metric is build right into the Wiener measure regularized 
phase-space path integral, which, along with the metric itself, is the 
centerpiece of metrical quantization.

Although we do not develop the subject further here, it is noteworthy 
that choosing a different geometry to support the Brownian motion 
generally leads to a qualitatively different quantization. For example, 
if the two-dimensional phase space has the geometry of a sphere of an 
appropriate radius, then metrical quantization leads not to canonical 
operators but rather to spin (or angular momentum) kinematical operators 
that obey the Lie algebra commutation relations of SU(2). On the other 
hand, for a phase space with the geometry of a space of constant 
negative curvature, metrical quantization leads to 
kinematical operators that are the generators of the Lie algebra for 
SU(1,1). Stated otherwise, the geometry of the chosen metric in 
postulate 1.~of metrical quantization---which then explicitly 
appears in the expression
defining the Wiener measure regularization in postulate 2.---actually 
determines the very nature of the kinematical operators in the metrical 
quantization procedure \cite{kla}.

\end{document}